\documentclass[%
 reprint,
 amsmath,amssymb,
 aps,
]{revtex4-2}

\usepackage{graphicx}
\usepackage{dcolumn}
\usepackage{bm}
\usepackage{enumitem}
\usepackage{xcolor} 
\usepackage{soul}
\definecolor{blizzardblue}{rgb}{0.67, 0.9, 0.93}

\renewcommand{\selectlanguage}[1]{}

\begin{document}

\preprint{APS/123-QED}

\title{Dual Quadrature Phasemeter for Space-Based Interferometry}

\author{Callum S. Sambridge}
\email{Callum.sambridge@anu.edu.au}
\author{Kirk McKenzie}%
\affiliation{%
 Centre for Gravitational Astrophysics, The Australian National University, Canberra ACT 2600, Australia
}%




\date{\today}

\begin{abstract}
This letter presents a {dual quadrature phasemeter}, an implementation of a phase-locked loop designed to track the phase of homodyne and heterodyne inter-satellite laser links. The dual quadrature phasemeters use dual quadrature optical detection to enable an alternate phase readout scheme that {operates with wider bandwidths and on signals with carrier frequency differences down to DC}. 
{Analytical modeling demonstrates that dual quadrature phasemeters overcome the bandwidth limitations of conventional phasemeters}. 
Numerical simulations {of} noise {linearity} tests found that the phase tracking error of the dual quadrature phasemeter is less than 10 microcycle/$\sqrt{\text{Hz}}$ in the presence of non-linear cyclic errors. The {dual quadrature phasemeter} {enables} exploration of a new optical configuration for retroreflector-based space geodesy missions with architecture similar to that implemented in GRACE-FO. One such alternate configuration is proposed that is capable of tracking satellite separation without requiring a frequency offset between local and incoming light, eliminating the need for optical frequency shifters.
\end{abstract}

\maketitle

\section{Introduction}

Phase locked loops (PLLs) are control loops that enable real time phase measurement of an oscillating signal. PLLs are used broadly across modern electronics, including in wireless and optical data communications, clock synchronization and frequency synthesis~\cite{hsieh_phase-locked_1996, stephens_phase-locked_2002}, largely due to their simplicity in implementation and versatility in application. PLLs have found further application in the field of optical metrology, where they are used to track phase changes in the interference between optical fields~\cite{shaddock_overview_2006} and phase-lock optical sources. The GRACE-FO mission is one such example, where inter-satellite laser links spanning 200~km are tracked with a phasemeter, a digitally implemented phase-locked loop, to measure changes in the Earth's ice mass and groundwater movement~\cite{landerer_extending_2020, bachman_flight_2017}. The phasemeter is uniquely suited for the abstraction of phase information from weak amplitude and high-frequency carrier signals on device, converting the mission data into a form that can be practically transferred to {E}arth~\cite{shaddock_overview_2006}. 

Phasemeters place design constraints on future Earth geodesy missions. {One} example is an alternate optical configuration proposed for the European Space Agencies (ESA) Next Generation Gravity Mission (NGGM), named the {retroreflector} configuration~\cite{nicklaus_laser_2020}, in which the optical payload of one satellite is replaced by a retroreflecting optic. This architecture simplifies optical design but brings with it several technical challenges. Double passing the optical link results in twice the divergence loss, making it challenging to meet received optical power requirements set by the phasemeter. Previous research undertaken by these authors, however, has showed that optical power margins can be relaxed for {retroreflector} configurations{ compared to {active transponder} configurations}, making these losses tolerable~\cite{sambridge_comparing_2024}. 
In the retroreflector configuration, science measurement is made by interfering the return light with local light from the same laser, resulting in a DC-centered interference frequency that varies with inter-satellite Doppler shifts ($\pm1$~megahertz). This signal is not suitable for standard phasemeter tracking, as the carrier frequency is outside the phasemeter's specified frequency input range (for GRACE-FO {4-16~MHz}~\cite{bachman_flight_2017}). When the carrier frequency of the tracked signal falls below the internal filter corner frequency of the PLL, up-shifted harmonics from the internal mixer dominate the phase output, corrupting measurement. Retroreflector designs accommodate this requirement by including a frequency shifter, making such designs incompatible with the current GRACE-FO optical bench~\cite{nicklaus_laser_2020}.

\begin{figure*}[ht]
    \centering
    \includegraphics[width = \linewidth]{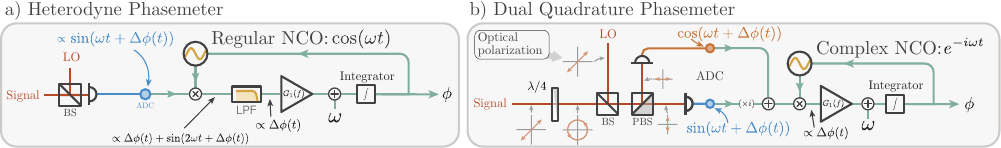}
    \caption{Block diagram schematic of heterodyne and dual quadrature phasemeter. a) heterodyne PLL detects the interference between two optical fields and mixes the digitized signal with a reference oscillator, consisting of a single numerically controlled oscillator (NCO). Mixing results in an output with up-shifted and down-shifted components. The higher frequency component is filtered and the downshifted component, which is proportional to the phase error between the reference and the input signal, is fed back to update the reference oscillator. b) Dual quadrature phasemeter combines the two input optical fields in a dual quadrature detection scheme that interrogates both quadratures of the interference signal. The quadrature inputs are digitized and combined to form a complex waveform before being mixed with a complex reference oscillator. Output of the complex mixing operation is only proportional to the phase error between the reference and the input{, removing the need for an in-loop second harmonic filter. The error signal is fed back to update the complex reference oscillator.}}
    \label{fig:PLL_figure}
\end{figure*}

While traditional PLLs are not suited for tracking the phase of DC-centered signals, feedback loop-based phase estimators for optical homodyne architectures have been developed in the field of optical communications. Taylor \cite{taylor_coherent_2004, taylor_phase_2009} presented a phase estimation loop that used a $90^\circ$~hybrid detection scheme to interrogate both quadratures of the incident optical field for demodulation of information encoded via phase shift keying. Taylor's scheme has since been proposed in ground-satellite optical communication schemes~\cite{paillier_space-ground_2020}. Sutton et al. \cite{sutton_digital_2013} applied a variant of the same concept in the context of homodyne digital interferometry, where a {homodyne PLL} used a CORDIC arc-tangent algorithm and feedback combined to measure displacement with a noise floor of 0.8pm/$\sqrt{\text{Hz}}$.

In this letter, we present {a {dual quadrature phasemeter}, a form of homodyne phase-locked loop} designed for phase metrology in homodyne and heterodyne systems. This implementation leverages dual quadrature optical detection to change the internal structure of the PLL. {Dual quadrature phasemeters} have two key advantages over heterodyne phasemeters --- They are able to track: 1) with a larger bandwidth than equivalent heterodyne PLLs
, and 2) signals with any carrier frequency between the positive and negative Nyquist frequencies of the capture rate. 
These benefits make them suitable for both homodyne and heterodyne detection schemes and enable tracking of higher frequency dynamics. An interferometric architecture for future Earth geodesy missions is proposed that leverages the tracking frequency range of the {dual quadrature phasemeter}, resulting in a retroreflector-based mission with a GRACE-FO-like optical bench. The {dual quadrature phasemeter} extends the class of optical signals that can be interrogated with real-time phase tracking techniques.

\section{Phase lock loop architecture}

\subsection{Heterodyne Phasemeter}

Fig \ref{fig:PLL_figure}.a). shows a block diagram schematic of a heterodyne phasemeter implemented with a phase-locked loop. Two optical fields are interfered and detected through photodetection. The output voltage signal is digitized, resulting in a periodic signal with angular frequency $\omega$ and phase information $ \Delta \phi(t)$. The signal is mixed with a digital reference oscillator at the same carrier frequency $\omega$. This operation results in an up-shifted component at the sum of the two mixed frequencies and a down-shifted component at their difference: 

\begin{align}
    \text{M}_\text{out}(t) &= \sin(\omega t + \Delta \phi(t)) \times \cos(\omega t)\\
    &= \sin(\Delta\phi(t)) + \sin(2\omega t + \Delta\phi (t)). \label{eqn:Mout_regular}
\end{align}
For small phase error, (\ref{eqn:Mout_regular}) is approximated as
\begin{align}
    \text{M}_\text{out}(t) &\approx \Delta\phi(t) + \sin(2\omega t + \Delta\phi (t)).
\end{align}

The up-shifted {component} at $2\omega$ is filtered by the in-loop second harmonic filter, leaving the phase error term, which is fed back to update the reference oscillator. With insufficient filtering, the {upshifted term is also fed back,} drive{ing the} reference oscillator and compromising measurement. As such, the carrier frequency of the tracked signal, $\omega$, must be sufficiently larger than the corner frequency of the in-loop second harmonic filter. The in-loop second harmonic filter cutoff also constrains the maximum stable bandwidth, as the total PLL latency is typically dominated by the delay of the in-loop second harmonic filter. These restrictions limit the scope of PLL applications in optical tracking to heterodyne sensing architectures with {bandwidths below the cutoff frequency of the in-loop second harmonic filter} and interference frequencies above.

\subsection{Dual Quadrature Phasemeter}
\label{sec:dualQuadPM}
Figure \ref{fig:PLL_figure}b) shows the block diagram schematic for a {dual quadrature phasemeter}. The interference between two optical fields is detected using a dual quadrature detection scheme~\cite{keem_removing_2004, garreis_90_1991, taylor_phase_2009, pietzsch_scattering_1989}. The resulting quadrature components of the detected signal are combined to form a complex input signal. This input signal is then mixed with a complex NCO, seeded at the negative of the input tone carrier frequency, $-\omega$, resulting in a mixer output of:
\begin{align}
    \text{M}_\text{out}(t) &= e^{i(\omega t + \Delta \phi(T))} \times e^{-i\omega t}\label{eqn:Mout_complex}\\
    &= \cos(\Delta\phi(t)) + i\sin(\Delta \phi(t))
\end{align}
which for small phase errors simplifies to
\begin{align}
    \text{M}_\text{out}(t) &\approx 1 + i\Delta \phi(t).
\end{align}
The imaginary component of the mixer output is passed through a controller and used to update the reference oscillator. As both components of mixing are complex, we are performing single sideband demodulation, which results in only a downshifted term proportional to the phase error between the input signal and reference oscillator. As such, the in-loop second harmonic filter is no longer required to ensure PLL linearity, removing the aforementioned restrictions it placed on the PLL with it. 

\subsection{Frequency Response Comparison}
\begin{figure}
    \centering
    \includegraphics[width = \linewidth]{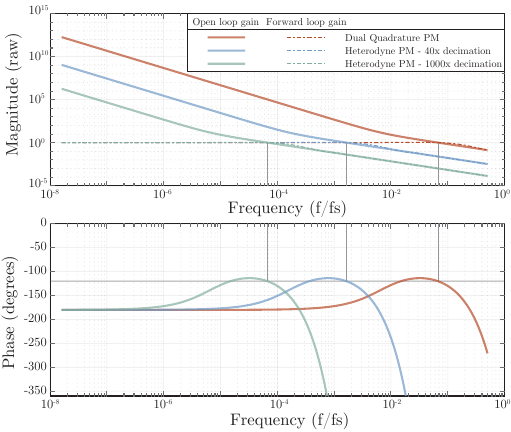}
    \caption{Bode plots of a dual quadrature phasemeter and two heterodyne phasemeters with internal filter corner frequencies of $f_s/40$ and $f_s/1000$. Top shows the magnitude, and the bottom shows the phase of frequency response. The controller gains of the phasemeter are tuned to maximize the unity gain frequency whilst maintaining a 60-degree phase margin. The maximum bandwidth of the two heterodyne phasemeters were constrained by the in-loop second harmonic filter delay. The dual quadrature phasemeter bandwidth was only limited by the propagation delay in the feedback loop, enabling it to track signals with higher frequency dynamics. }
    \label{fig:BODE_plot}
\end{figure}

Fig~\ref{fig:BODE_plot}. shows a Bode plot for a {dual quadrature phasemeter} (red) and two heterodyne phasemeters, with in-loop second harmonic filter cutoffs of $f_s/40$ (yellow) and of $f_s/1000$ (green). Frequency is given as a proportion of the digital clock rate of the PLLs, $f_s$. The phasemeter's open loop gains were modelled as

\begin{equation}
    H_{\tau}(s) = e^{-s \tau}\left( \frac{2\pi f_{\text{ug}}}{s} + 0.1\times\left(\frac{2\pi  f_{\text{ug}}}{s}\right)^2 \right)
\end{equation}

where{ $s = i 2 \pi f$ is the Laplace parameter, }$f_{\text{ug}}$ is the unity gain frequency of the {phasemeter}, and $\tau$ is the propagation delay in the phasemeter feedback loop. For {a phasemeter} with an in-loop second harmonic filter implemented as a cascade integrator-comb filter, the shortest possible time delay achievable by the loop is $\tau =  (N+1)/f_s$, where N is the decimation rate of the in-loop second harmonic filter (also the ratio between the in-loop second harmonic filter cutoff and the PLL update rate, $f_s$). This analysis assumes that, outside of the delay imposed by the in-loop second harmonic filter of the heterodyne phasemeter, the total propagation delay in the phasemeter feedback loop is limited to a single clock cycle. This generalization is made to emphasise the decoupling of the in-loop delay from the feedback loop components. 

The controllers of each {phasemeter} were tuned to maximize the unity gain frequency whilst maintaining a 60-degree phase margin. The unity gain frequency of the {dual quadrature phasemeter} was only limited by the update rate of the phasemeter (which in an idealized case is $\tau = 1/f_s$), whereas the maximum unity gain frequencies of the two heterodyne PLLs are limited by the delay imposed by their in-loop second harmonic filters. For the heterodyne PLL, the in-loop second harmonic filter also defines the minimum trackable frequency of the PLL. As such, there is a trade-off between how low a frequency the heterodyne PLL can track and how large a bandwidth it can track with. This limitation is not present in the {dual quadrature phasemeter}. Higher bandwidths enable the dual quadrature phasemeter to track signals with higher frequency dynamics, extending the class of signals it can track over the heterodyne phasemeter.

\subsection{Practical implementation}
One drawback of {dual quadrature phasemeters} over heterodyne phasemeters is their susceptibility to cyclic errors{: errors} caused by DC offsets or differences in the relative amplitudes or phase of the in-phase and quadrature signal components~\cite{keem_removing_2004}. Cyclic errors insert unsuppressed up-shifted mixing components into the feedback signal, which results in non-linear errors in phase measurement. Extensive literature exists on the correction of cyclic errors, as they are present in all quadrature detection measurement schemes~\cite{keem_removing_2004, eom_dynamic_2001, hu_compensation_2015}. Earth geodesy missions require the contribution of phasemeter non-linear tracking error to be below{ 1~nm/$\sqrt{\text{Hz}}$}~\cite{bachman_flight_2017}. As such, the following numerical simulations explore the performance of the dual quadrature phasemeter in the presence of static cyclic errors both in isolation and when implemented with a cyclic correction algorithm based on that presented in~\cite{hu_compensation_2015}.






\section{Linearity test}
A phasemeter linearity test used by Shaddock et al. in~\cite{shaddock_overview_2006}, named the three noise test, was used to explore the phase tracking performance of the {dual quadrature phasemeter}. The three noise test uses three phasemeters to track numerically generated laser frequency noise and combines the {phasemeter measurements} to remove the input signals, leaving only the non-common non-linear tracking errors. Cyclic errors in the {dual quadrature phasemeter} cause {harmonics} at multiples of the carrier frequency to be modulated onto the internally generated reference, mixing any signals at these frequencies into the phasemeter's measurement. When the carrier frequency of the signal is greater than the bandwidth of the phasemeter, any mixing harmonics caused by cyclic errors are suppressed by the frequency response of the phasemeter. As such, cyclic-induced non-linear errors are worse when the carrier frequency of the tracked signal is within the phasemeter bandwidth range around DC. 

\begin{figure}[t]
    \centering
    \includegraphics[width = \linewidth]{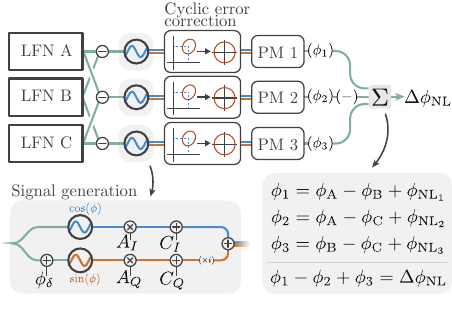}
    \caption{Phasemeter three noise test performed with three dual quadrature phasemeters. Laser frequency noise of $10\text{~Hz}/\sqrt{\text{Hz}} \times 1/f$ was numerically generated in three separate instances before being combined to form three linearly independent signals. The three signals drove numerically controlled oscillators where cyclic errors were inserted in the form of DC offsets, amplitude mismatches and phase errors. The complex output signals were fed through a cyclic error correction procedure from~\cite{hu_compensation_2015} to suppress the impact of the inserted cyclic errors. The resulting signals were tracked with three separate dual quadrature phasemeters, each with a unity gain frequency of 10~kHz. The outputs of the three phasemeters were combined to determine the total non-linear tracking error present.}
    \label{fig:CPM_3NT}
\end{figure}

Fig. \ref{fig:CPM_3NT} illustrates a block diagram schematic of the phasemeter three noise test. Laser frequency noise was simulated as frequency noise with a {$10\text{~Hz}/\sqrt{\text{Hz}} \times 1 \text{Hz}/f$} amplitude spectral density. The laser frequency noise was simulated with deliberately low amplitude to keep the tracked beatnotes within the 10~kHz phasemeter bandwidth of DC for the majority of each test, increasing the impact of non-linear errors on measurement. 
The carrier frequencies of the three signals {were} varied between simulations to explore performance in two cases: with the initial beatnote frequency at 1~MHz and with the initial frequency at 0~Hz (DC).
The first case is an idealized case, where the carrier frequency is far from DC. The second case is a more challenging tracking environment, notably a worse case than Earth geodesy missions, in which inter-satellite induced Doppler shifts pull the carrier frequencies away from DC ({between $\pm 1$~MHz for GRACE-FO}). 


The simulated laser frequency noise was used to generate two quadrature components of the input signal. In signal generation, a constant randomized phase offset (within $\pm 0.1$ cycles), amplitude scaling (within $\pm 10\%$), and DC offsets (within $\pm 10\%$ of amplitude) were included to simulate static cyclic errors. After generation, each signal {was} passed through a cyclic correction algorithm based on~\cite{hu_compensation_2015}. After correction, each complex signal was tracked with an independent dual quadrature phasemeter. The phase outputs of the three phasemeters were linearly combined to remove the simulated laser phase noise, leaving only non-linear tracking errors and the tracking noise floor of the phasemeters. Three cases were explored with the simulation:
\begin{enumerate}[label = (\roman*)]
    \item \textbf{DC cyclic} - DC-centered tracking with cyclic errors and no corrections applied; exploring worst-case tracking error.
    \item \textbf{DC corrected} - DC-centered tracking with cyclic errors and corrections applied, determining the impact of the corrections on improving measurement.
    \item \textbf{1~MHz tracking} - 1~MHz centered tracking with no cyclic errors to provide a baseline for performance in ideal conditions. 
\end{enumerate}

\begin{figure}[h!]
    \centering
    \includegraphics[width = \linewidth]{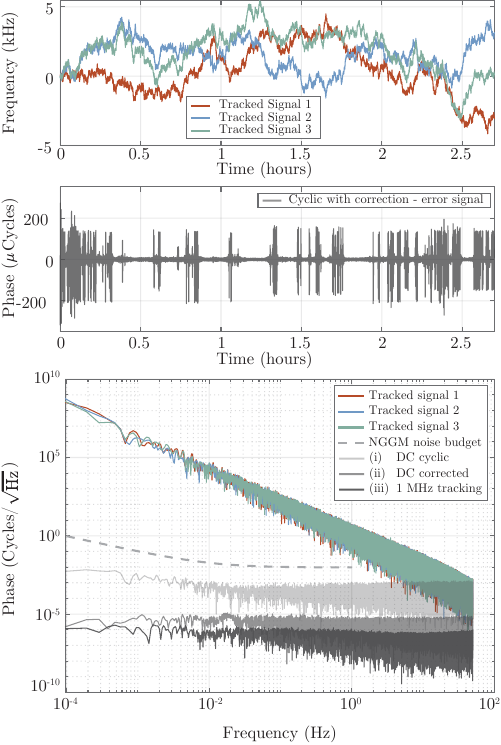}
    \caption{Results from the phasemeter three noise test performed on the dual quadrature phasemeter. Top: tracked frequency signals by the three phasemeters. Middle: non-linear tracking error for a DC-centered signal with cyclic errors present and cyclic correction active. Bottom: phase amplitude spectral density of the three tracked signals and tracking error for the three cases explored: DC-centered signal with cyclic error (light grey), DC-centered signal with cyclic correction applied (dark grey) and 1~MHz centered ideal tracked signal (black). The noise level of the DC-centered signal with cyclic corrections present was $30$~dB below the NGGM noise requirement~\cite{nicklaus_towards_2022}, demonstrating the initial feasibility for a dual quadrature phasemeter-based measurement scheme.}
    \label{fig:3NT_results}
\end{figure}

Fig \ref{fig:3NT_results}. displays the results of the three noise tests. Top shows the frequencies of the three signals tracked by the three phasemeters for a single measurement. Middle shows the non-linear phase error (three noise test output) for the {DC corrected} measurement of the signals displayed in the top plot. Bottom shows the phase amplitude spectral densities of the three tracked phase signals as well as the residual errors of three different cases: light gray --- \textbf{DC cyclic}, dark gray --- \textbf{DC corrected}, and black --- \textbf{1~MHz tracking}. 

{Tracking performance varied between the three cases explored.} The 1~MHz tracking measurement demonstrates the noise floor of the phasemeter in ideal cyclic conditions is 1~microcycle/$\sqrt{\text{Hz}}$. This performance is equivalent to the noise level observed by Shaddock et al.~\cite{shaddock_overview_2006} in evaluating a phasemeter for LISA.
Introducing cyclic errors into the tracked signal raised the phasemeter noise floor by 3 orders of magnitude relative to the 1~MHz tracking case, and further rolling up below 10~mHz. With correction, however, the impact of these cyclic errors could be reduced to below 10~microcycle/$\sqrt{\text{Hz}}$ {(10~pm/$\sqrt{\text{Hz}}$ with a 1064~nm wavelength carrier)}, well below the 1~nm/$\sqrt{\text{Hz}}$ phasemeter performance requirement for geodesy missions. The tracking error ASD of the corrected case is largely dominated by spurs of noise that occur when one of the tracked signals was within 100Hz of DC. This suggests that with either improved cyclic error correction or in an environment where less time is spent within 100~Hz of the DC, phasemeter performance could be further improved. In the case of GRACE-FO, the Doppler-induced frequency shift is within $\pm 100$~Hz for {four} 1~second instances each 90~minute orbit.



\section{Implementation in earth geodesy mission}

\begin{figure}
    \centering
    \includegraphics[width = \linewidth]{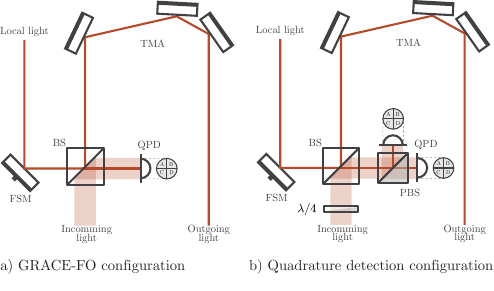}
    \caption{Optical configuration for a geodesy mission using the dual quadrature phasemeter. a) Simplified GRACE-FO optical bench structure. Local light is steered with a fast steering mirror (FSM) before being interfered with incoming light on a 90/10 beam splitter. The transmitted port is split on a 50/50 beam splitter (BS) and detected on two quadrature photodetectors (QPD) for redundancy. The reflected port is redirected through a triple mirror assembly and transmitted to the far satellite. b) Alternate optical configuration that enables dual quadrature coherent detection and application of the dual quadrature phasemeter. Incoming light is passed through a quarter wave plate before interference with local light on the same 90/10 beam splitter. The beam splitter is replaced with a polarizing beam splitter (PBS) and two quadrant photodetectors to interrogate the in-phase and quadrature components of the incident optical field.}
    \label{fig:GRACE-FO_change}
\end{figure}

Figure \ref{fig:GRACE-FO_change} a) and b) illustrate a simplified GRACE-FO-type optical mission architecture and an alternate architecture that enables the application of the {dual quadrature phasemeter}{ respectively}. This alternative architecture allows a GRACE-FO-like LRI to function without a frequency offset between the local and incoming light, thereby enabling retroreflector-based schemes to operate without including a frequency shifter. The authors note that they do not see any benefit to using this optical architecture in a {transponder} scheme, such as that implemented in GRACE-FO. 

In the GRACE-FO optical configuration{, Figure} \ref{fig:GRACE-FO_change} a), an incoming weak field is interfered with a bright reference field on an uneven beam splitter and detected using a single quadrant photo-diode. On the other output of the beam splitter, the bright field is redirected by the triple mirror Assembly (TMA) and transmitted to the far satellite. In the second configuration{, Figure }\ref{fig:GRACE-FO_change} b), the incoming weak field is passed through a quarter wave-plate, circularly polarizing the light. This delays one polarization component by $\lambda /4 $ relative to the other, enabling dual quadrature detection~\cite{kazovsky_all-fiber_1987}. The weak field is then interfered with the bright local light (which needs to be vertically polarized at 45 degrees relative to the polarizing beam splitters) on a 90-10 beam splitter (90\% reflected). The transmission path for the local light is further split through a polarizing beam splitter and detected with two quadrant photo-diodes. At this detection point, two quadrant photo-diodes measure the I and Q components of the signal. The second output (reflected path for the LO) is redirected by the triple mirror assembly and transmitted to the second satellite. This alternate configuration enables quadrature coherent detection whilst leaving {much} of the preexisting GRACE-FO optical configuration unchanged. 

This design facilitates the exploration of missions that feature a simplified second satellite that uses a passive retroreflector. Asymmetric retroreflector-based inter-satellite interferometer, as investigated in~\cite{nicklaus_laser_2020, amata_system_2019}, can achieve full system redundancy, a significant strength over the current GRACE-FO architecture. However, less explored asymmetric designs may present opportunities for further reductions in cost and mission complexity. For instance, the second satellite could operate without an onboard accelerometer, employing the accelerometer transplant method demonstrated by GRACE-FO~\cite{bandikova_grace_2019}. Furthermore, such a mission could leverage the lower optical power and frequency dynamic tracking requirements of retroreflector-based systems, as shown in~\cite{sambridge_comparing_2024}, to eliminate the need for higher power lasers and enable the application of C-band lasers.

While these results suggest the feasibility of the {dual quadrature phasemeter} for space-based interferometric missions, there are several other factors that need to be considered. Without added frequency shifters, the operational frequency band of the mission would be a DC-centered range of $\pm$ the maximum Doppler-induced frequency shift. This imposes noise requirements on laser relative intensity noise (RIN), photodetector dark noise and ADC noise in this DC-centered frequency band. {While a space-based interferometric geodesy mission has not previously been designed to operate in this band, meeting the GRACE-FO RIN and electronic noise requirements~\cite{nicklaus_laser_2020} in a DC-centered band is possible with consumer-grade components}. The authors find no compelling reason to suggest that this system would fail, thereby justifying further investigation

\section{Conclusions}
This letter presents a phase-locked loop designed specifically for implementation in homodyne and heterodyne optical metrology, named here the {dual quadrature phasemeter}. By redesigning the structure of a PLL around a coherent detection scheme, {dual quadrature phasemeters} can track optical signals with higher frequency dynamics and with relaxed carrier frequency requirements. Analytical modelling showed that the technique overcomes bandwidth limitations present in heterodyne phasemeters. Numerical simulations demonstrated that, when paired with correction techniques, the tracking error present in a {dual quadrature phasemeter} in the presence of cyclic errors was below 10~microcycle$/\sqrt{\text{Hz}}$ for a simulated DC-centered signal. This technique enables a retroreflector-based sensing architecture for inter-satellite laser links that operates without requiring an optical frequency shifter, making it more compatible with a GRACE-FO-like optical bench. 

\section*{Funding}
This research was supported by the Australian Research Council Centre of Excellence for Gravitational Wave Discovery (OzGrav, CE230100016), and the Australian Research Council Centre of Excellence for Engineered Quantum Systems (EQUS, CE170100009).


\section*{Disclosures}
The authors declare no conflicts of interest. 

\bibliography{zotero_references}

\end{document}